\documentclass{article}
\usepackage[utf8]{inputenc}
\usepackage[square,sort,comma,numbers]{natbib}
\usepackage{graphicx}
\usepackage{epstopdf}
\usepackage{inputenc}
\usepackage{amsmath}
\usepackage{amssymb}
\usepackage{listings}
\usepackage[bottom=0.8in, top=0.8in]{geometry}
\usepackage[
  separate-uncertainty = true,
  multi-part-units = repeat
]{siunitx}

\usepackage{xcolor}
\usepackage{appendix}
\usepackage{mathtools}
\usepackage{caption}
\usepackage{subcaption}
\usepackage{float}
\usepackage{hyperref}
\usepackage{algorithm}
\usepackage{tikz}

\usepackage{listings}
\usepackage{xcolor}
\usetikzlibrary{shapes,arrows,positioning,calc}
\usepackage{sectsty}
\allsectionsfont{\textsf}
\usepackage[affil-it]{authblk}
\usepackage{algpseudocode}
\title{NanoFrame: A web-based DNA wireframe design tool for 3D structures }
\author{Samson Petrosyan, Grigory Tikhomirov}
\affil{Electrical Engineering and Computer Science, 
    University of California, Berkeley}
 
\colorlet{punct}{red!60!black}
\definecolor{background}{HTML}{EEEEEE}
\definecolor{delim}{RGB}{20,105,176}
\colorlet{numb}{magenta!60!black}
 
 \lstdefinelanguage{json}{
    basicstyle=\normalfont\ttfamily,
    numbers=left,
    numberstyle=\scriptsize,
    stepnumber=1,
    numbersep=8pt,
    showstringspaces=false,
    breaklines=true,
    frame=lines,
    backgroundcolor=\color{background},
    literate=
     *{0}{{{\color{numb}0}}}{1}
      {1}{{{\color{numb}1}}}{1}
      {2}{{{\color{numb}2}}}{1}
      {3}{{{\color{numb}3}}}{1}
      {4}{{{\color{numb}4}}}{1}
      {5}{{{\color{numb}5}}}{1}
      {6}{{{\color{numb}6}}}{1}
      {7}{{{\color{numb}7}}}{1}
      {8}{{{\color{numb}8}}}{1}
      {9}{{{\color{numb}9}}}{1}
      {:}{{{\color{punct}{:}}}}{1}
      {,}{{{\color{punct}{,}}}}{1}
      {\{}{{{\color{delim}{\{}}}}{1}
      {\}}{{{\color{delim}{\}}}}}{1}
      {[}{{{\color{delim}{[}}}}{1}
      {]}{{{\color{delim}{]}}}}{1},
}
\begin{document}

\maketitle
\vspace{\fill}
\begin{abstract}
The rapid development of the DNA nanotechnology field has been facilitated by advances in CAD software. However, as more complex concepts arose, the lag between the needs and software capabilities appeared. Further derailed by manual installation and software incompatibility across different platforms and often tedious library management issues, the software has become hard-to-use for many.

Here we present NanoFrame, a web-based DNA wireframe design tool for making 3D nanostructures from a single scaffold. Within this software, we devised algorithms for DNA routing, staple breaking, and wireframe cage opening which, while modeled for cuboid structure, can be generalized to a variety of platonic and Archimedean shapes. In addition, NanoFrame provides a platform for editing auto-generated staple sequences and saving work online.

\end{abstract}
\vspace{\fill}
\subsection*{Supplementary Material}
\textit{Website: } \href{http://nanoframe.org}{\texttt{nanoframe.org}}
\\
\textit{Repository: } \href{https://github.com/tilabberkeley/nanoframe}{\texttt{github.com/tilabberkeley/nanoframe}}
\\
\textit{Tutorials: } \href{http://www.nanoframe.org/miscellaneous\#tutorial}{\texttt{nanoframe.org/miscellaneous\#tutorials}}
\vspace{\fill}

\newpage
\section{Introduction}
In his pioneering work Ned Seeman\citep{1} showed that DNA can be programmed to form shapes with nanometer scale precision by reengineering the natural replication junction. The field of DNA nanotechnology picked up pace with Paul Rothemund’s invention of DNA Origami technique \citep{2}, where a long viral DNA (scaffold) is folded into any desired shape with the help of many synthetic DNA strands (staples). Since the inception of DNA origami, multiple software have been built to aid the design process of DNA nanostructures. Douglas et al. released the widely used caDNAno \citep{3} tool for design of densely-packed DNA origami nanostructures.  Soon after, other software packages such as DAEDALUS, TALOS, and PERDIX \citep{4, 5, 6} and plug-ins like Adenita \citep{7} followed, enabling wireframe designs. Although these software packages are powerful tools for DNA wireframe design with different edges (double crossover (DX) for DAEDALUS and two-helix bundle (2HB) for TALOS), they require manual installation and are not available across different operating systems and language versions. These factors present a substantial barrier of entry to researchers who do not have the necessary technical background. Therefore, these software tools are not often utilized to the best of their capabilities.

Here we present NanoFrame, an open-source web-based wireframe design application that embodies the notion of software as a service (SaaS) taken from software engineering discipline \citep{8}. The creation of NanoFrame as SaaS was driven by the immense overhead cost of software installation and version management and the immobility of these software development methods. NanoFrame fully utilizes continuous integration and deployment, wherein updates to the software are delivered instantaneously to the users. Within a few seconds of deployment a user may start working with an updated version of the software. NanoFrame is a graphical user interface (GUI) based tool and it automates the process of designing DNA nanostructures to a great extent. Getting staples sequences can be as simple as clicking four buttons. While the software provides a full autopilot for the design pipeline, it also enables users to manually edit routing and placement of strands. NanoFrame is greatly inspired by scadnano \citep{9}, the only current fully online software. Scadnano is designed for closely-packed DNA origami, while NanoFrame is designed for wireframe structures.

The main reason we focus on wireframe structures is that they cover larger surface area compared to conventional Rothemund’s DNA origami with the same amount of DNA (see Section 2.3 for more details). This is important for embedding molecular recognition capabilities properties of DNA into non-DNA materials to guide their assembly, that is our overarching goal. This molecular information embedding will enable several nanophotonics applications \citep{10} as well as a aid drug delivery to diseased cells \citep{20}.  Thus, unlike its counterpart software \texttt{small}, that requires multiple origamis \citep{11}, NanoFrame designs wireframe cages from a single scaffold, covering the surface of similarly sized 50 nm cubic nanoparticles. Wireframe’s drawback is lower density of surface or edge extensions, while its advantages are lower cost, expected higher yield (per surface area unit), and stability is broader range of salt concentration. 
The first version of NanoFrame only supports cuboid structures, due to cuboids' immediate need in nanophotonics and nanoplasmonics \citep{12}. We show the path to extend NanoFrame algorithms to arbitrary shapes.

\section{Methods}
Creative ways have been proposed for finding routing by modeling 3D shapes as graphs comprised of a set of vertices and edges \citep{13, 14}. This abstraction enables the use of concepts and algorithms from graph theory. One such example was proposed by Benson et al. for finding routing by solving the associated `Chinese postman tour' problem. This and other previous works have focused on finding a routing by modeling the entire object as a single graph. Here we propose an alternative approach of finding a routing by only working with 2D planes of a shape.
\subsection{Routing}
The planes of all cuboid shapes are defined on a standard square grid, where positions start at $(0, 0)$ and go until $(s, s)$ where $s$ signifies the number of segments or the number of stripes added one. As the topology does not change and is merely a matter of scaling the coordinates, the dimensions are left to be readjusted on the front end. Within the grid, each integer position set except the corners are transformed into vertices in a graph with edges connecting top, bottom, left, and right integer neighbors (where permissible). Vertices that are divisible by $s$ are marked as outgoers and the rest are classified as singletons (Figure 1). 
When a cube is dissected, it can be observed that the edges can be connected in a way that results in the dissected cube being a planar graph. The four-color theorem \citep{15} proposes that no more than four colors are necessary to color any planar graph. By extending this idea of color to routing, we claim that no more than four unique plane routings are needed to construct a routing of any cuboid structure. With four unique planes, there are a total of $4^6 = 4096$ possible combinations that result in cuboid structures. Brute force search for the cuboid with a single loop - that corresponds to a scaffold - can be computed in constant time.

Finding four unique routings led us to a non-deterministic algorithm. Specifically, here we introduce randomized depth-first search. In each iteration of the algorithm, two random outgoer vertices are selected - denoted as $s$ and $t$ - and a modified depth-first search is run with a predefined decision rule. For cuboids, this decision rule states that if a horizontal edge was taken in the last step, a vertical step must be taken next, and vice versa. 
\begin{figure}[H]
    \centering
    \includegraphics[scale=0.5]{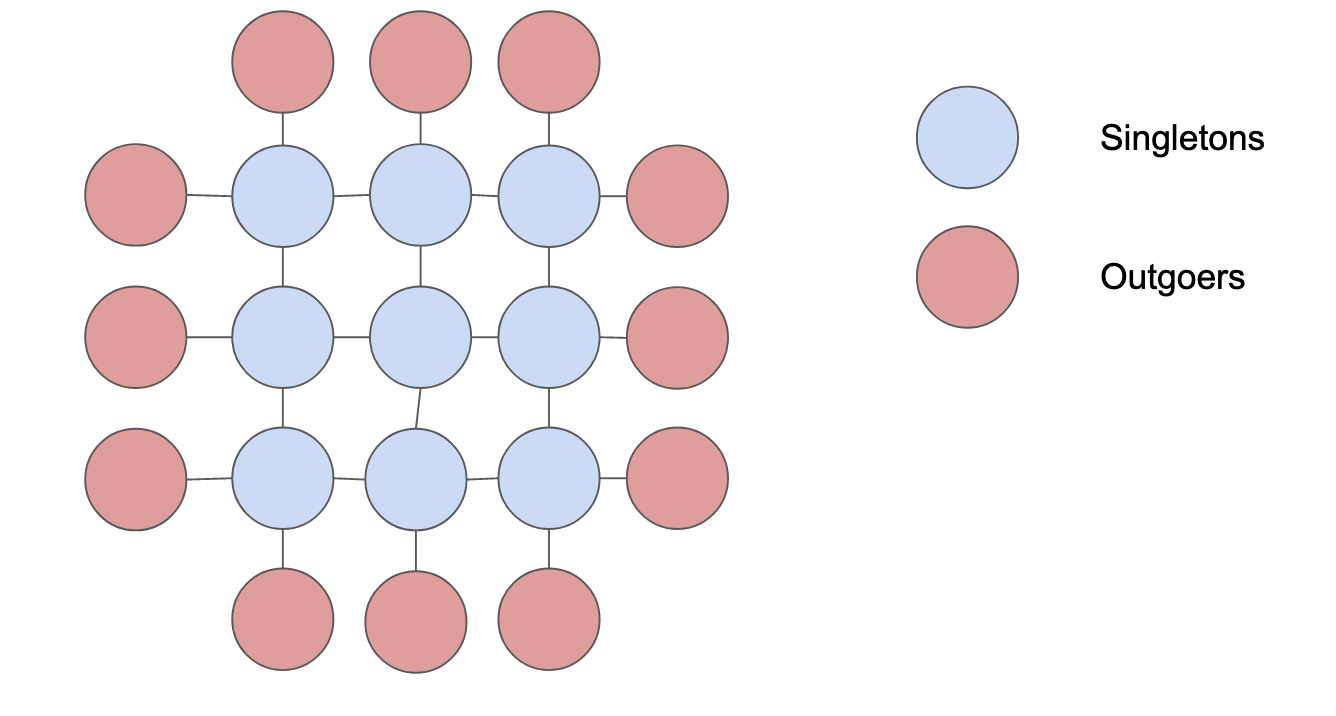}
    \caption{Graph Modeling of the 3D Object with 3 stripes or 4 segments}
    \label{fig:my_label}
\end{figure}
If a path is found from $s$ to $t$, then a list of edges taken is returned, otherwise an empty list (Figure 2). After each iteration of DFS, the graph is augmented by removing all the taken edges and working with the residual graph. The algorithm halts when there are no more remaining vertices and edges. Figure 3 shows the cuboid routings found for different dimensions by using the routings found in planes through algorithms 1-3.

\begin{figure}[H]
     \centering
     \begin{subfigure}[b]{0.4\textwidth}
         \centering
         \includegraphics[width=\textwidth]{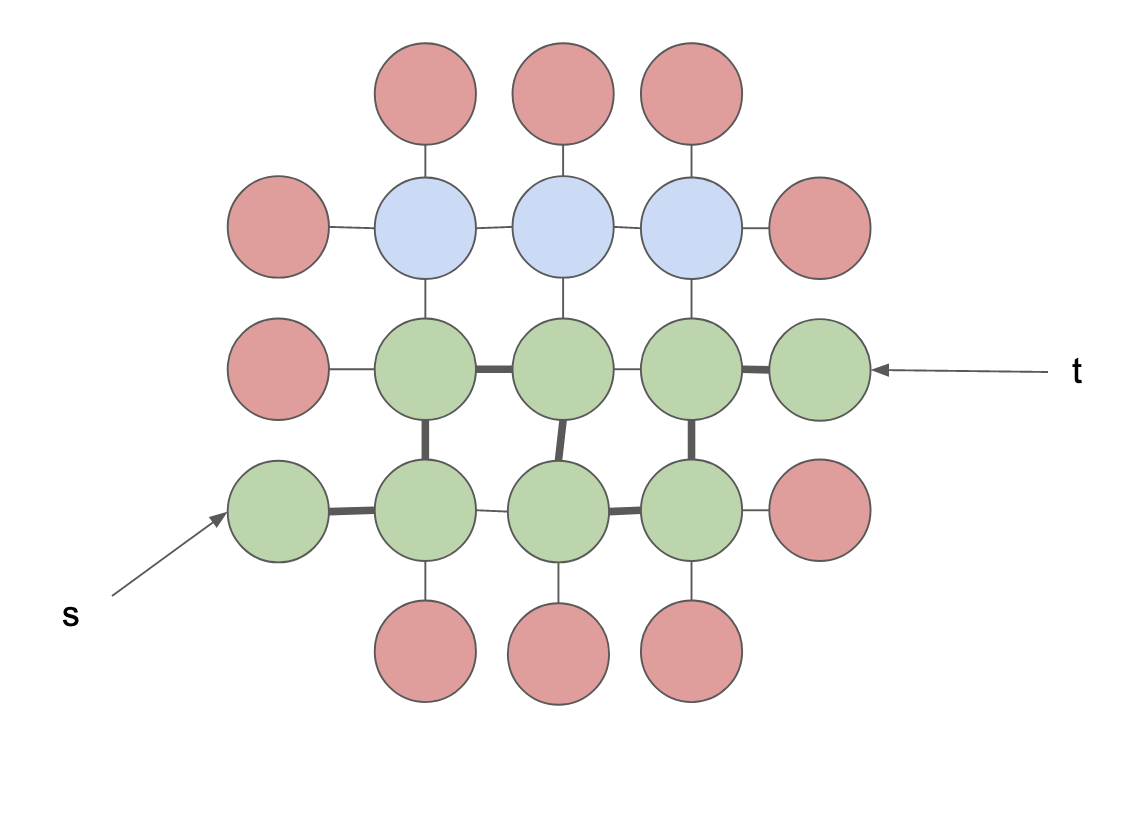}
         \caption{}
         \label{fig:y equals x}
     \end{subfigure}
     $\rightarrow$
     \begin{subfigure}[b]{0.4\textwidth}
         \centering
         \includegraphics[width=\textwidth]{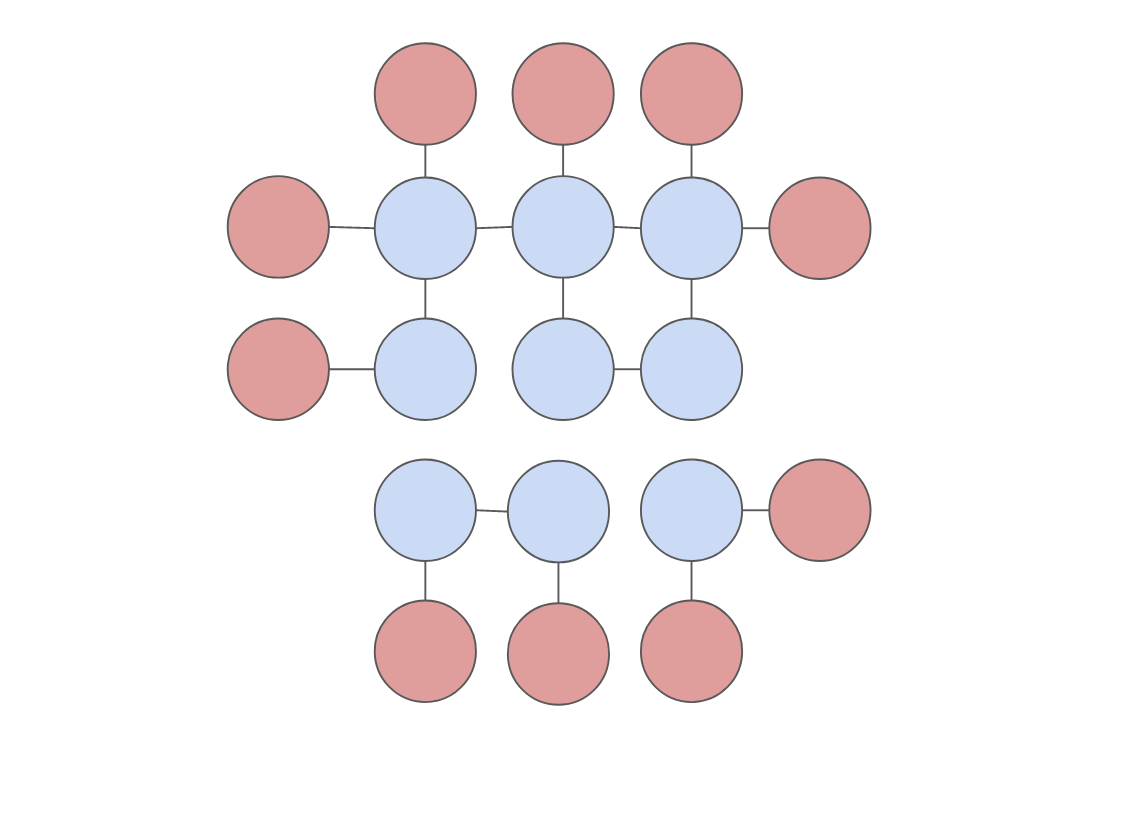}
         \caption{}
         \label{fig:three sin x}
     \end{subfigure}
     \caption{First iteration of Randomized DFS. (a) A path is found from $s$ to $t$. (b) The underlying graph is augmented by removing the edges of the path and outgoers.}
\end{figure}

\begin{algorithm}[H]
\caption{find\_general\_plane\_routing()}\label{alg:cap}
\begin{algorithmic}

\State $\texttt{outgoers} \gets \texttt{find\_outgoers}$
\State $\texttt{total\_outgoers} \gets \texttt{outgoers.length}$
\State $\texttt{taken\_outgoers} \gets \texttt{[]}$
\State $\texttt{taken\_edges} \gets \texttt{[]}$
\State $\texttt{sets} \gets \texttt{[]}$
\While{$\texttt{taken\_outgoers.length != total\_outgoers}$}
    \State $\texttt{s} \gets \texttt{outgoers[rand(0..(outgoers.length - 1))]}$
    \State $\texttt{outgoers.delete(s)}$
    \State $\texttt{t} \gets \texttt{outgoers[rand(0..(outgoers.length - 1))]}$
    \State $\texttt{dfs\_edges} \gets \texttt{dfs(s, t, taken\_edges)}$

    \If {$\texttt{dfs\_edges != []}$}
        \State $\texttt{outgoers.delete(t)}$
        \State $\texttt{taken\_outgoers << s}$
        \State $\texttt{taken\_outgoers << t}$
        \State $\texttt{taken\_edges.concat(dfs\_edges)}$
        \State $\texttt{new\_set = Set.new(s)}$
        \State $\texttt{new\_set.add\_node(t)}$
        \While{$\texttt{dfs\_edges has edge e}$}
            \State{$\texttt{new\_set.add\_edge(e)}$}
        \EndWhile
        \State{$\texttt{sets << new\_set}$}
    \Else
        \State{$\texttt{outgoers << s}$}
    \EndIf
\EndWhile

\If {$\texttt{taken\_edges.length != @edges.length}$}
    \State {$\texttt{return find\_general\_plane\_routing()}$}
\Else
    \State {$\texttt{return sets}$}
\EndIf
\end{algorithmic}
\end{algorithm}

\begin{algorithm}[H]
\begin{algorithmic}
\caption{dfs(k, t, edges)}\label{alg:cap}
\State{$\texttt{visited = \{\}}$}
\State{$\texttt{prev} \gets \texttt{k.x \% @segments == 0 ? 0 : 1}$}
        
\While {$\texttt{@vertices has vertex v}$}
    \State{$\texttt{visited[v.hash]} \gets \texttt{[]}$}
\EndWhile
\State {$\texttt{visited} \gets \texttt{explore(k, t, prev, edges, visited)}$}
\State {$\texttt{return visited[t.hash]}$}
\end{algorithmic}
\end{algorithm}

\begin{algorithm}[H]
\begin{algorithmic}
\caption{explore(k, t, prev, edges, visited)}\label{alg:cap}

\State{$\texttt{neighbors} \gets \texttt{find\_neighbors(k, prev, edges)}$}
\If {$\texttt{neighbors.length == 0}$}
    \State {$\texttt{return []}$}
\EndIf
\While {$\texttt{neighbors has neighbor}$}
    \State {$\texttt{new\_edge = Edge.new(k, neighbor)}$}
    \State {$\texttt{edges = find\_and\_remove\_edge(edges, new\_edge)}$}

    \If {$\texttt{visited[neighbor.hash] == []}$}
        \While {$\texttt{visited[k.hash] has path}$}
                   \State {$\texttt{visited[neighbor.hash] << p}$}
        \EndWhile
    \EndIf
    \State {$\texttt{explore(neighbor, t, (prev - 1).abs(), edges, visited)}$}
\EndWhile
\State {$\texttt{return visited}$}
\end{algorithmic}
\end{algorithm}

\begin{figure}[H]
     \centering
     \begin{subfigure}[b]{0.3\textwidth}
         \centering
         \includegraphics[width=\textwidth]{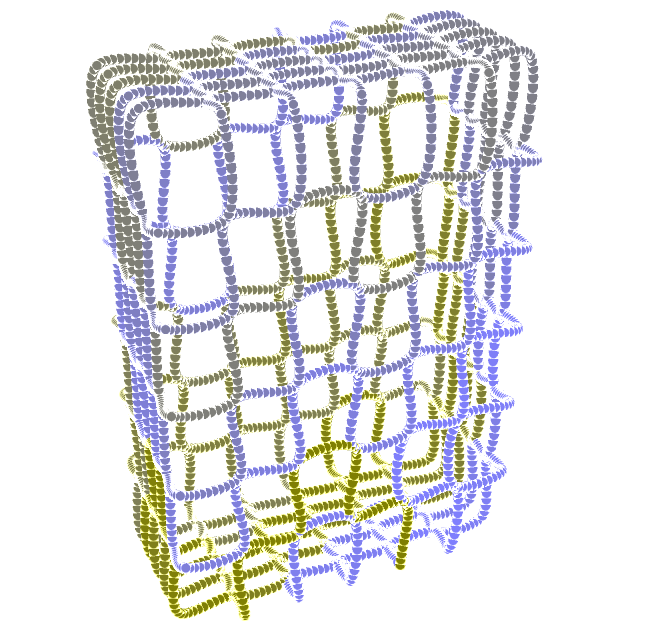}
         \caption{}
         \label{fig:y equals x}
     \end{subfigure}
     \hfill
     \begin{subfigure}[b]{0.3\textwidth}
         \centering
         \includegraphics[width=\textwidth]{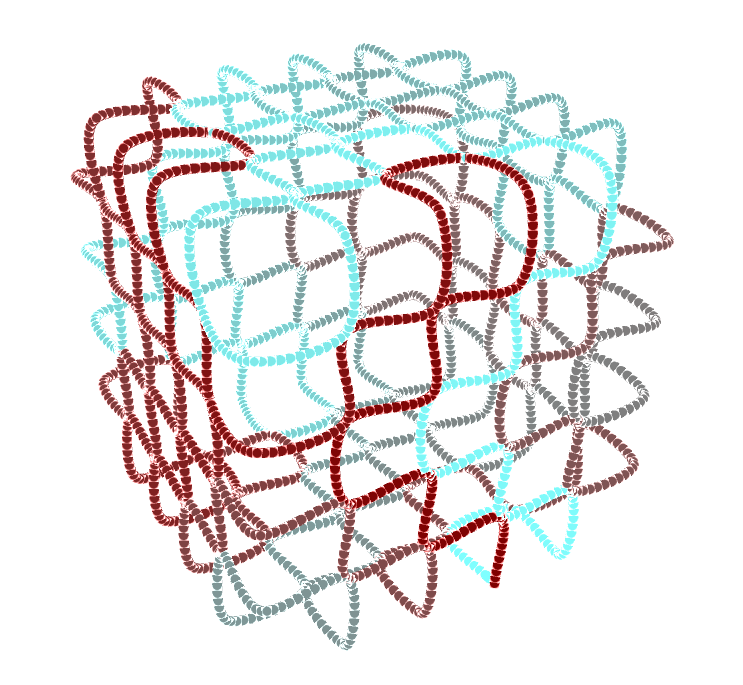}
         \caption{}
         \label{fig:three sin x}
     \end{subfigure}
     \hfill
     \begin{subfigure}[b]{0.3\textwidth}
         \centering
         \includegraphics[width=\textwidth]{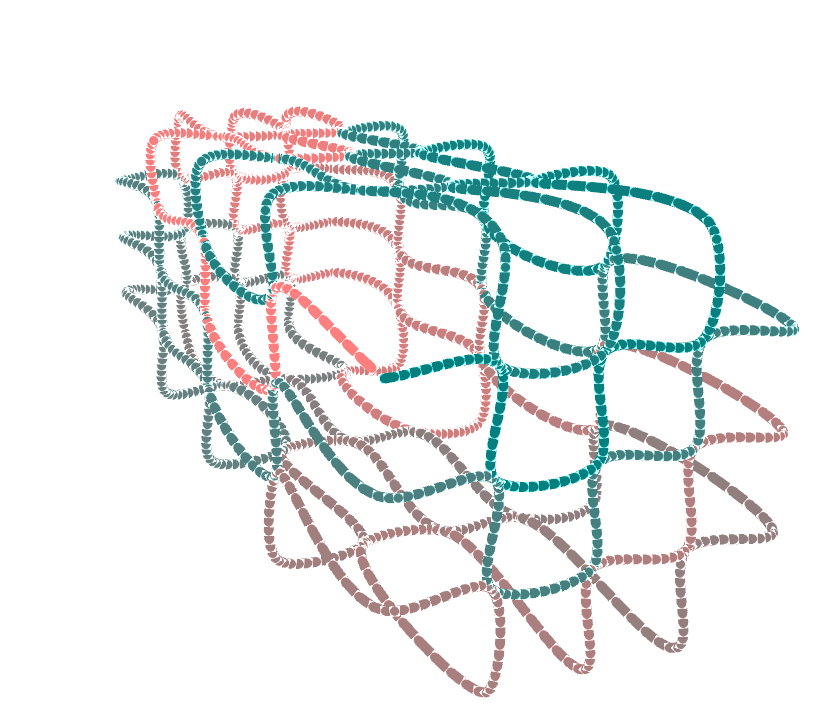}
         \caption{}
         \label{fig:five over x}
     \end{subfigure}
        \caption{(a) 40nm x 60nm x 20nm by 5 stripes cuboid. (b) 50nm x 50nm x 50nm by 4 stripes cuboid. (c) 40nm x 40nm x 120nm by 3 stripes cuboid. }
        \label{fig:three graphs}
\end{figure}

\subsection{Staple Categorization and Breaking}
For any cuboid object, three distinct staple categories are defined: reflections, refractions, and extensions. Reflections connect two orthogonal parts of the scaffold that are non-sequential edges. Refractions are projections onto the neighboring plane and are the staples that cross from one plane onto the other. Extension strands that have their own four subcategories fill the gap between reflections and refractions.

We provide a staple breaking algorithm that is run for each of the planes of the cuboid structure. Staple breaking is a linear programming algorithm, which comes with preprocessing and postprocessing stages (Figure 4). The input $x$ maps from the dimensions and stripes of the object. In preprocessing, a set of linear constraints are developed that will feed into the linear program solver in the next step. Two booleans values corresponding to width and height are computed for whether it is possible to generate staples without extensions. If staple sequence exists such that no constraints are violated for both with and without extensions, a pop-up window is displayed to the user to select preferred staple generation, otherwise the corresponding constraints are removed. The output $y$ - the set of linear constraints - is then inputted in a linear program solver like simplex \citep{16} which returns the results $Q(y)$. Finally, these results are then rounded to the nearest integer values in the postprocessing stage alongside breaking staples into smaller pieces for long extensions and for staples that open the wireframe (see Section 2.3 for more details). The final result is a vector $P(x)$ which corresponds to the lengths of staples. 
\\
\begin{figure}[H]
    \centering
    \begin{tikzpicture}[node distance=1cm]
\tikzstyle{parent} = [rectangle, minimum width=10cm, minimum height=3cm, draw=black, fill=white]
\tikzstyle{startstop} = [rectangle, minimum width=3cm, text width=20mm, minimum height=1cm,text centered, draw=black, fill=white]
\tikzstyle{alg} = [rectangle, text width=40mm, minimum height=1cm,text centered, draw=black, fill=white]
\tikzstyle{arrow} = [thick,->,>=stealth]
\node (input) [] {$x$};
\node (preprocess) [startstop, right=of input] {Preprocess};
\node (algorithm) [alg, right=of preprocess] {Simplex with predefined objective function and $y$ constraints};
\node (postprocess) [startstop, right=of algorithm] {Postprocess};
\node (output) [right=of postprocess] {$P(x)$};

\draw [arrow] (input) -- node[anchor=south] {} (preprocess);
\draw [arrow] (preprocess) -- node[anchor=south] {$y$} (algorithm);
\draw [arrow] (algorithm) -- node[anchor=south] {$Q(y)$} (postprocess);
\draw [arrow] (postprocess) -- node[anchor=south] {} (output);
\end{tikzpicture}
    \caption{Staple Breaking Flow}
    \label{fig:my_label}
\end{figure}
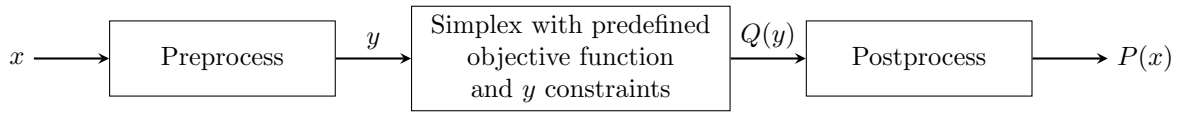

In the most general form, where all types of staples are possible the linear program consists of 6 variables and 13 constraints. The $x = \begin{bmatrix}s & w & h & d\end{bmatrix}^T$ is the input vector which carries the values of stripes, segment width and height, and the dimension of the scaffold accordingly. 
\[\boxed{\begin{array}{rrclcl}
\displaystyle \max_{x, y, z_1, z_2, z_3, z_4} & \multicolumn{5}{l}{4sx + 2s^2y+2sz_{1}+2sz_{2} + (s^2-s)z_3 + (s^2-s)z_4} \quad \quad \\
\textrm{s.t.} & x, y, z_1, z_2, z_3, z_4 \geq 20 \quad \\
&\displaystyle  x, y \leq 60 \quad \\
& \frac{x}{2} + \frac{y}{2} + z_1 = w \quad \\
& \frac{x}{2} + \frac{y}{2} + z_{2} = h \quad\\
& y + z_{5} = w \quad\\
& y + z_{6} = h \quad \\
& 4sx + 2s^2y+2sz_{1}+2sz_{2} + (s^2-s)z_3 + (s^2-s)z_4 < d \quad \\
\end{array}}\]
The output vector $P(x) = \begin{bmatrix}x & y & z_1 & z_2 & z_3 & z_4\end{bmatrix}^T$ carries the lengths of each type of staple strand of the plane.

\subsection{Opening Wireframes}
To put nanoparticles inside a wireframe object, we develop a procedure that identifies the strongest connected components in any three planes starting from one-third of the scaffold size up to two-third.  Strongest connected component for a wireframe with given scaffold cross-section $c$ (where $c\in[\frac{1}{3}d\dots\frac{2}{3}d]$, $d$= length of scaffold) is defined as the maximum number of edges in any three given planes divided by the total number of edges for cross-section $c$. Though the range of values is by default between $\frac{1}{3}d\dots\frac{2}{3}d$, it could be manually adjusted by the user. The connection between the two strongest connected components then serves as the switch for opening the wireframe object. As routing may not be unique, regenerating routing can yield better strongest connected components which would, in turn, increase the stability of the wireframe object in an open form.

Once the points connecting the strongest connected components are identified, the binding staple sequences in the neighboring region are weakened by making them shorter in the postprocessing part of the staple breaking algorithm. The specific lengths of the new staples will be dictated by the dimensions of the object. With shorter staple strands, the wireframe object can be opened and closed by increasing and decreasing the temperature in a mixer. Once a nanoparticle is entrapped in one of the strongest connected components, the wireframe object is closed (Figure 5).

\begin{figure}[H]
     \centering
     \begin{subfigure}[b]{0.45\textwidth}
         \centering
         \includegraphics[width=\textwidth]{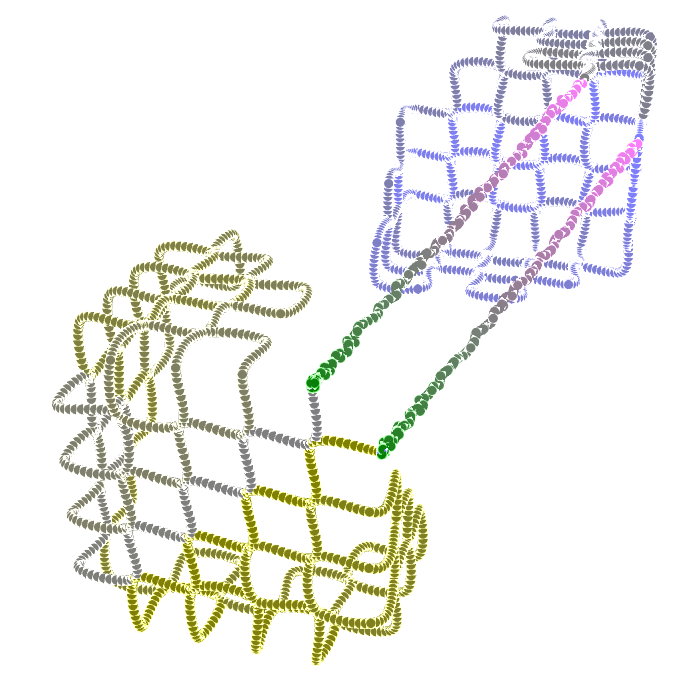}
         \caption{}
         \label{fig:y equals x}
     \end{subfigure}
     \hfill
     \begin{subfigure}[b]{0.45\textwidth}
         \centering
         \includegraphics[width=\textwidth]{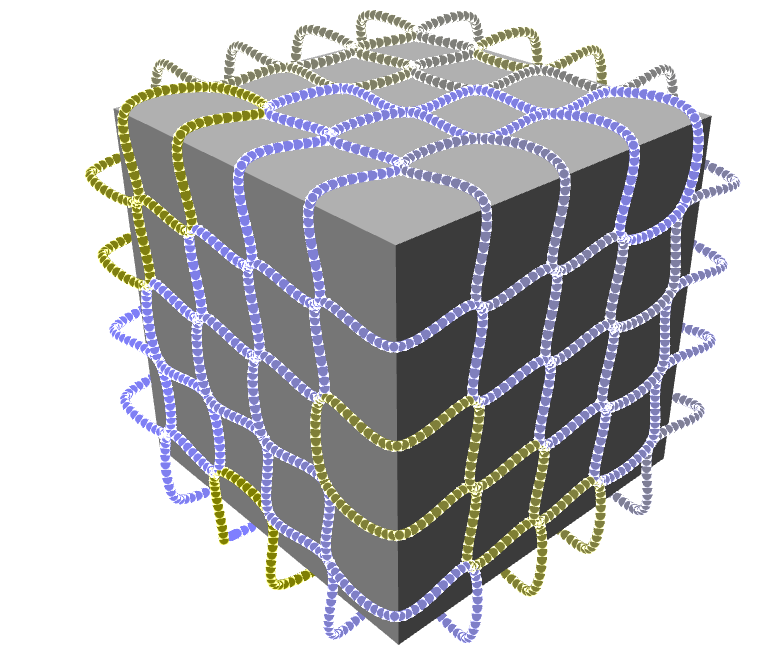}
         \caption{}
         \label{fig:three sin x}
     \end{subfigure}
        \caption{(a) Open cuboid wireframe with artificially extended connection edges. (b) Nanoparticle enclosed in a DNA wireframe. }
        \label{fig:three graphs}
\end{figure}
\subsection{Manual Editing}
Although staple breaking aims to alleviate the humdrum of manually editing staple sequences, for some tasks it is often necessary to make manual edits to the automatically generated staples. NanoFrame provides tools where users can navigate through the plane routings while at the same time seeing the 3D object in the background. The edit toolbox enables almost any desired modification of the staples and scaffold, much like scadnano but with the enhanced ability to work with non-parallel helices.
\begin{figure}[H]
    \centering
\includegraphics[scale=0.3]{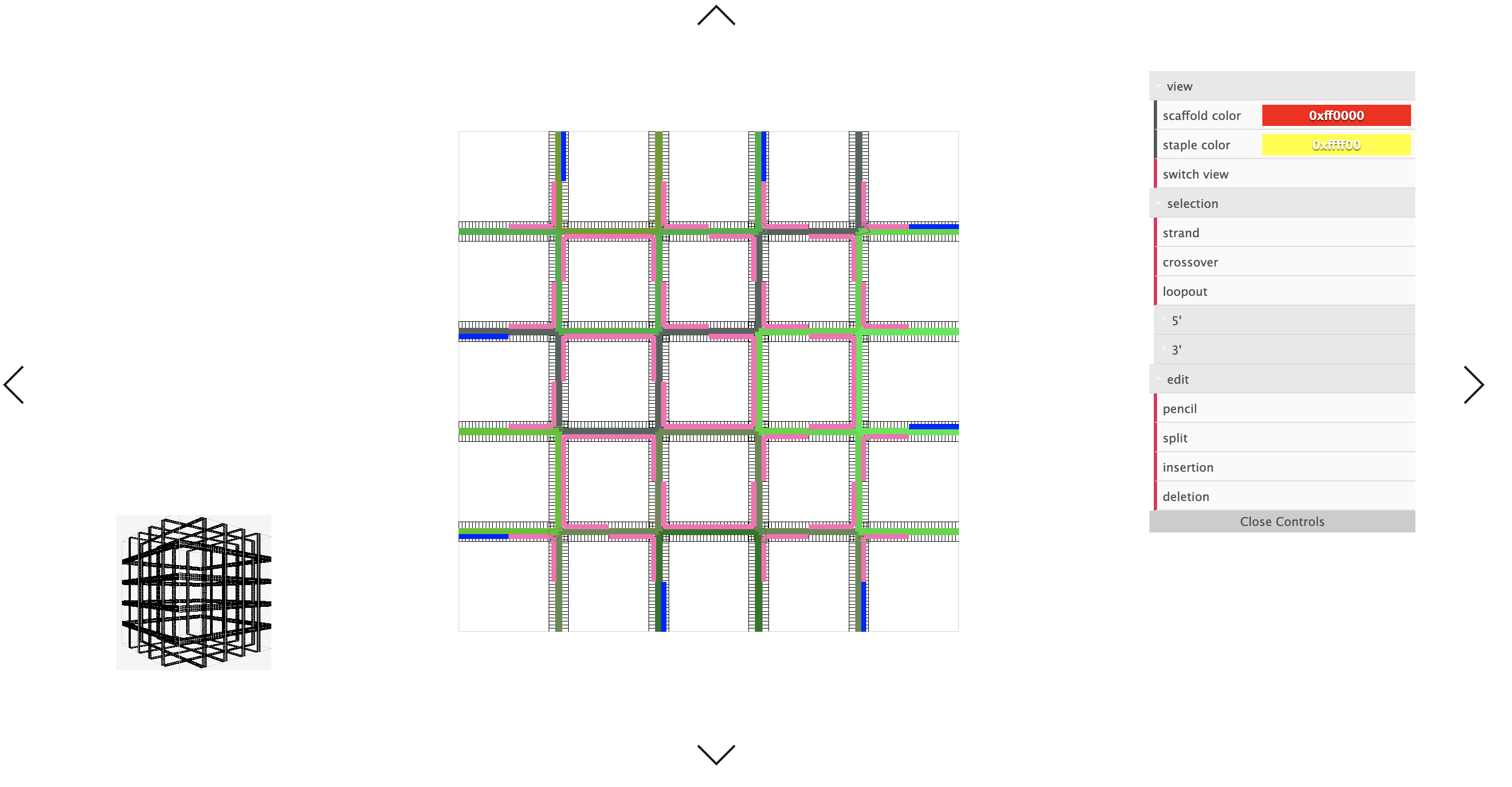}
    \caption{Manual edit for 50nm x 50nm x 50nm by 4 stripes cuboid}
    \label{fig:my_label}
\end{figure}
\subsection{Atomic Synthesization}
As an integral part of the software, NanoFrame is able to realize the design into widely used formats like \texttt{pdb} and \texttt{oxview}. Figure 5 shows the atomic synthesization of the shapes created in Figure 3. Downloads are available with and without staple strands attached.
\begin{figure}[H]
     \centering
     \begin{subfigure}[b]{0.3\textwidth}
         \centering
         \includegraphics[width=\textwidth]{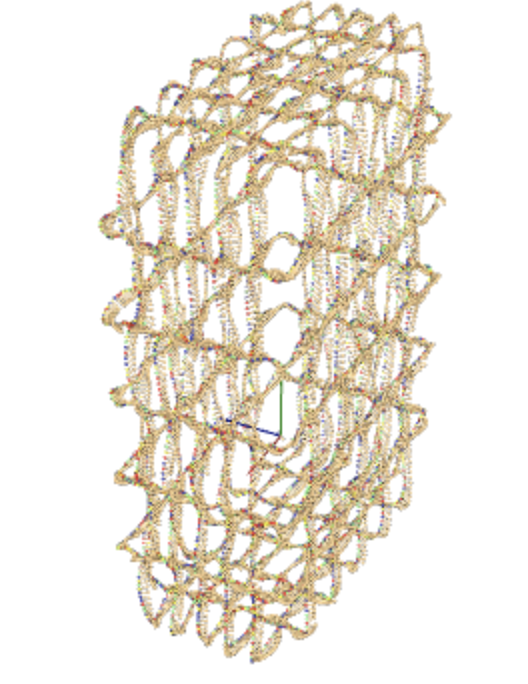}
         \caption{}
         \label{fig:y equals x}
     \end{subfigure}
     \hfill
     \begin{subfigure}[b]{0.3\textwidth}
         \centering
         \includegraphics[width=\textwidth]{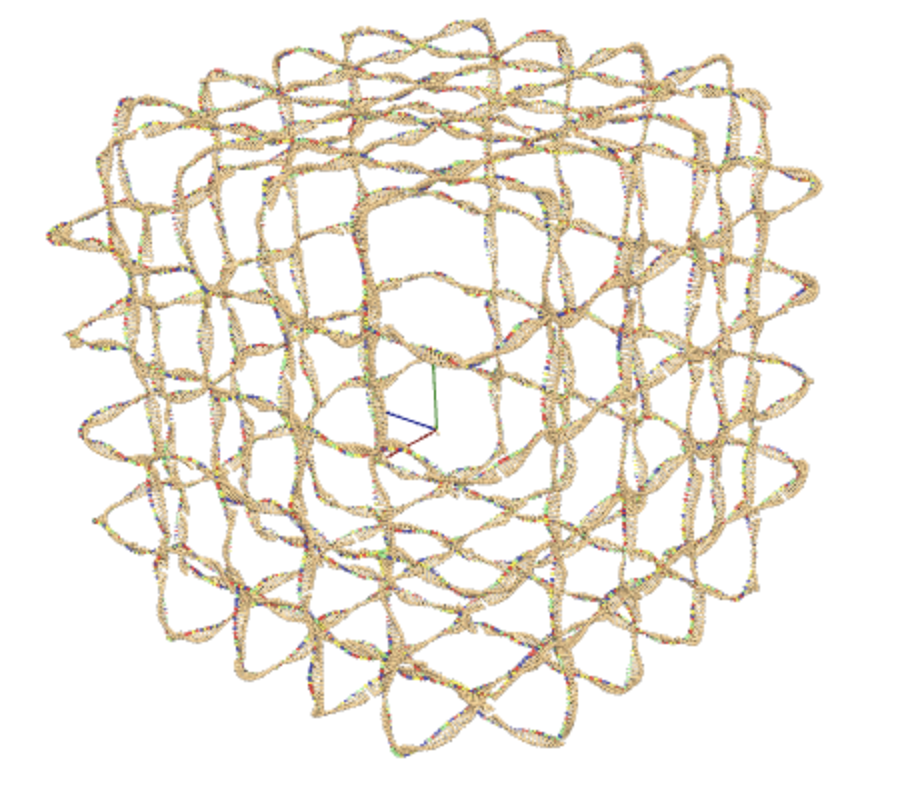}
         \caption{}
         \label{fig:three sin x}
     \end{subfigure}
     \hfill
     \begin{subfigure}[b]{0.3\textwidth}
         \centering
         \includegraphics[width=\textwidth]{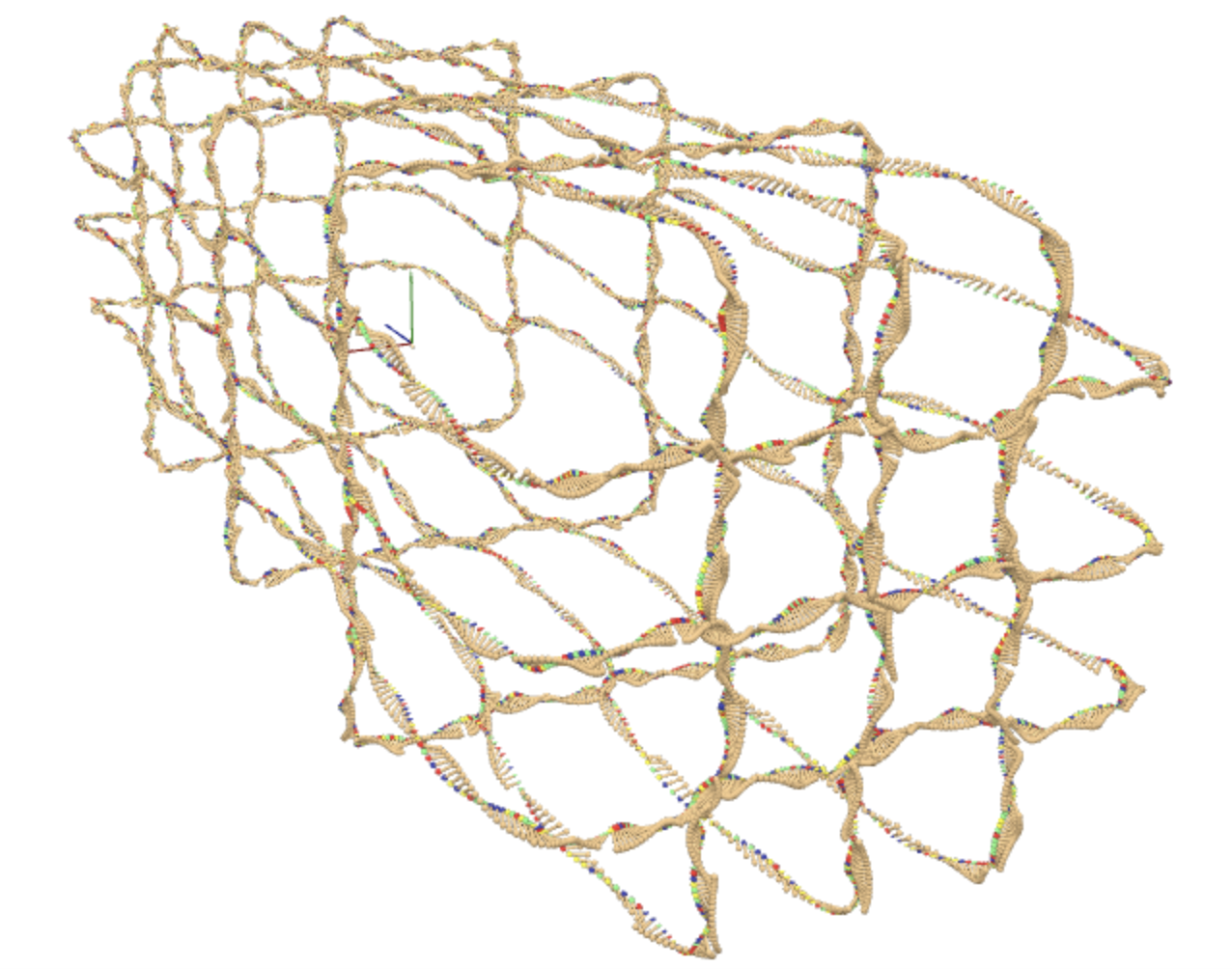}
         \caption{}
         \label{fig:five over x}
     \end{subfigure}
        \caption{(a) 40nm x 60nm x 20nm by 5 stripes cuboid. (b) 50nm x 50nm x 50nm by 4 stripes cuboid. (c) 40nm x 40nm x 120nm by 3 stripes cuboid. }
        \label{fig:three graphs}
\end{figure}
Although only two DNA file formats are available through NanoFrame, the conversion between the provided formats and the desired ones can be achieved by tacoxDNA \citep{17} which is also embedded in NanoFrame.
\section{Server}
NanoFrame is a Ruby on Rails application, and as a result, it provides a server and database for storing design implementations. Fundamentally, all files are generated from NanoFrame native \texttt{nfr} format which itself is JSON. As both \texttt{pdb} and \texttt{oxdna} are memory intensive files, these do not get stored in the database, and instead it is the routing \texttt{nfr} and the staple sequence \texttt{csv} files that are stored online. \texttt{pdb} and \texttt{oxdna} are still feasible to download after saving work, but these are computed on-demand. 
An \texttt{nfr} file contains plane objects which themselves contain an array of sets. Each of these sets represents a routing across a plane and is therefore constructed of edges and vertices. A sample \texttt{nfr} file is shown below.

\begin{lstlisting}[language=json,  numbers=none, basicstyle=\tiny]
{
     sets: [
         // start plane 1
         {
             edges: [
                 {
                     v1: {
                         x: 0,
                         y: 0,
                         z: 0
                     }, 
                     v2: {
                         x: 0,
                         y: 15,
                         z: 0
                     }
                 },
                 {
                     v1: {
                         x: 0,
                         y: 15,
                         z: 0
                     }, 
                     v2: {
                         x: 15,
                         y: 15,
                         z: 0
                     }
                 }
             ]
         },
         {
             edges: [
                 {
                     v1: {
                         x: -15,
                         y: 15,
                         z: 0
                     }, 
                     v2: {
                         x: 0,
                         y: 15,
                         z: 0
                     }
                 },
                 {
                     v1: {
                         x: 0,
                         y: 15,
                         z: 0
                     }, 
                     v2: {
                         x: 0,
                         y: 30,
                         z: 0
                     }
                 }
             ]
         }
     // end plane 1
     /* start plane 2
      ...
        end plane 2 */
     ] 
}
\end{lstlisting}
Users can save their work in private or public mode. Designs made public will be displayed in the feed of all users. Private designs can involve collaboration by either inviting users through NanoFrame username or email. Each user gets a history page which lists all the designs they have created. Those that were not saved simply have the name of the shape and its dimensions, whereas saved work also includes the routing \texttt{nfr} file and the staples \texttt{csv} file.
\begin{figure}[H]
    \centering
    \includegraphics[scale=0.2]{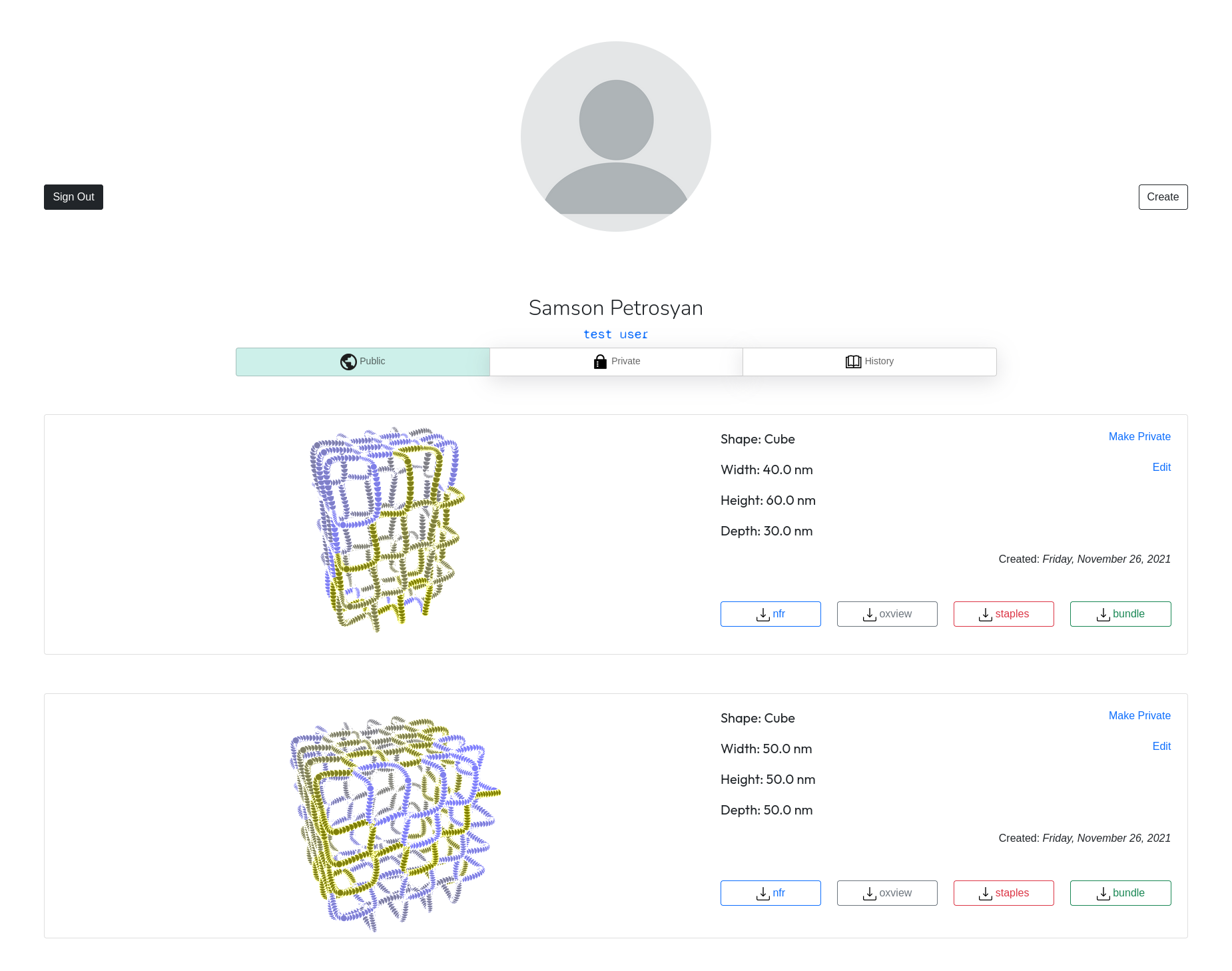}
    \caption{User profile in NanoFrame homepage}
    \label{fig:my_label}
\end{figure}
\section{Future Work}
In this paper, we have drawn the blueprint of a web-based approach of 3D wireframe nanostructure design through DNA. NanoFrame solves one of the bottlenecks of the current state of software in the field by modeling itself as a software as a service, a paradigm that has proven to be both efficient and resourceful in software engineering \citep{18}. NanoFrame yearns to reduce the difficulty of using the software while at the same time providing a rigorous set of tools that go on par with non-SaaS applications. At its current state, NanoFrame can only design and synthesize cuboid objects, but though these are certainly useful, the ability to generate routing for other platonic and Archimedean shapes has shown promising applications \citep{13, 19} and will further be developed. The next step of the evolution process of NanoFrame is to enable the design of more complex 3D objects that could also be user-built, either through uploading CAD files like \texttt{ply} or through sketching on canvas in NanoFrame. 

The staple breaking algorithm provides a generic approach to obtaining DNA staple strands through the use of three types of staple categories. We believe that this categorization can translate to other 3D objects. Strongest connected component algorithm also provides a blueprint for opening arbitrary wireframe shapes. Though for cuboids the optimal number of strongest connected components is two, for other shapes the number may vary. Future work will involve improving the algorithm such that the optimal number of strongest connected components is extracted from the wireframe shape. The routing algorithm can be modified to include a directional vector which would be generated through a shape such that the edge selection in the randomized depth-first search stems from this vector. We believe that these algorithms - alongside others developed in this project - can greatly streamline the design of wireframe objects using DNA.

\end{document}